\documentclass[prd,aps,nofootinbib,amsmath,amsfonts,amssymb]{revtex4}
\usepackage[pdftex,colorlinks]{hyperref}

\def \BE {\begin{equation}}
\def \EE {\end{equation}}
\def \BEA { \begin{eqnarray}}
\def \EEA {\end{eqnarray}}

\def \d {\mathrm{d}}
\def \D {\mathrm{D}}
\def \bn {\mbox{\boldmath{$n$}}}
\def \bk {\mbox{\boldmath{$k$}}}
\def \bl {\mbox{\boldmath{$\ell$}}}
\def \bm {\mbox{\boldmath{$m$}}}

\begin{document}

\title{Type III and N solutions to quadratic gravity}

\author{Tom\' a\v s M\'alek}
\email{malek@math.cas.cz}
\affiliation{Institute of Theoretical Physics, Faculty of Mathematics and Physics, Charles University,  
V Hole\v sovi\v ck\' ach 2, 180 00 Prague 8, Czech Republic}

\author{Vojt\v ech Pravda}
\email{pravda@math.cas.cz}
\affiliation{Institute of Mathematics, Academy of Sciences of the Czech Republic, \v Zitn\' a 25, 115 67 Prague 1, Czech Republic}

\date{\today}

\begin{abstract}
We study exact vacuum solutions to quadratic gravity (QG) of the Weyl types N and III. 
We show that in an arbitrary dimension all Einstein spacetimes of the Weyl type N  with an appropriately chosen
effective cosmological constant $\Lambda$ are exact solutions to QG and we refer
to explicitly known metrics within this class.

For type III Einstein spacetimes, an additional constraint follows from the field equations of QG 
and examples of spacetimes obeying such constraint are given. However,  type III pp-waves
do not satisfy this constraint and thus do not solve QG.    

For type N, we also study a wider class of spacetimes admitting a pure radiation term
in the Ricci tensor. In contrast to the Einstein case, the field equations of generic QG determine
optical properties of the geometry and restrict such exact solutions to  the Kundt class.
We provide examples of these metrics.
\end{abstract}

%\pacs{}

\maketitle

\section{Introduction}

In perturbative quantum gravity, corrections have to be added to the Einstein action.
Demanding coordinate invariance, these corrections should consist of various curvature invariants.
One important class of such modified gravities is quadratic gravity (QG) \cite{Deser2002},
whose action contains general quadratic terms in curvature
\begin{equation}
  S = \int \d^n x \sqrt{-g} \bigg( \frac{1}{\kappa} \left( R - 2 \Lambda_0 \right)
      + \alpha R^2 + \beta R^2_{ab}
      + \gamma \left(R^2_{abcd} - 4 R^2_{ab} + R^2 \right) \bigg).
  \label{QG:action}
\end{equation} 

Varying \eqref{QG:action} with respect to the metric leads to vacuum quadratic gravity field equations \cite{GulluTekin2009}
\begin{eqnarray}
  \frac{1}{\kappa} \left( R_{ab} - \frac{1}{2} R g_{ab} + \Lambda_0 g_{ab} \right)
  + 2 \alpha R \left( R_{ab} - \frac{1}{4} R g_{ab} \right)
  + \left( 2 \alpha + \beta \right)\left( g_{ab} \Box - \nabla_a \nabla_b \right) R \nonumber \\
  + 2 \gamma \bigg( R R_{ab} - 2 R_{acbd} R^{cd}
  + R_{acde} R_{b}^{\phantom{b}cde} - 2 R_{ac} R_{b}^{\phantom{b}c}
  - \frac{1}{4} g_{ab} \left( R^2_{cdef} - 4 R^2_{cd} + R^2 \right) \bigg) \nonumber \\
  + \beta \Box \left( R_{ab} - \frac{1}{2} R g_{ab} \right)
  + 2 \beta \left( R_{acbd} - \frac{1}{4} g_{ab} R_{cd} \right) R^{cd} = 0.
  \label{QG:fieldeqns}
\end{eqnarray}

These field equations are very complex and a direct approach to finding exact solutions seems to be hopeless.
Nevertheless a few exact solutions, such as four-dimensional plane wave \cite{Madsen1990}, are known.
Very recently, $n$-dimensional AdS waves solving quadratic gravity were also found in \cite{Gullu2011} using Kerr--Schild ansatz. 

In recent years, an algebraic classification of the Weyl tensor generalizing the four-dimensional Petrov classification to arbitrary dimension was developed \cite{Coleyetal04} (see also \cite{Reall2011} for  an introductory  review).
Such classification is based on the existence of preferred null directions --- Weyl aligned null directions (WANDs) and their multiplicity.
Spacetimes not admitting WANDs are of type G and spacetimes admitting WANDs of multiplicity 1, 2, 3 and 4 are of the principal Weyl type I, II, III and N, respectively.
Note that both exact solutions of QG mentioned above are of the Weyl type N.
Furthermore, for types N and III the Weyl part of the Kretschmann invariant $R_{abcd} R^{abcd}$ appearing in the action identically vanishes which leads to a considerable simplification of the field equations \eqref{QG:fieldeqns}. We will thus seek for exact solutions of QG of the Weyl types N and III.  

In section \ref{sec_einst}, we study Einstein spacetimes as exact solutions of QG for dimension $n>4$.\footnote{In four
dimensions, where the Gauss--Bonnet term does not contribute to the field equations, all Einstein spacetimes
are exact solutions of QG \cite{LuPope2011}.}
We show that all type N Einstein spacetimes with appropriately chosen
effective cosmological constant $\Lambda$  are exact solutions of QG and we refer to explicit examples
of such spacetimes given in the literature.

For type III Einstein spacetimes, vanishing of a quantity $\tilde \Psi$ \eqref{tPsi} plays a key role.
Namely, it follows that  type III Einstein spacetimes (with appropriately chosen
effective cosmological constant $\Lambda$) are exact solutions of QG if and only if $\tilde \Psi=0$.
We give examples of type III spacetimes with both $\tilde \Psi=0$ and $\tilde \Psi \neq 0$.
Interestingly, the case $\tilde \Psi \neq 0$ contains also all type III Ricci-flat pp-waves and therefore these pp-waves are {\it not} exact solutions of QG. We also compare the class of exact solutions of QG with other related classes of spacetimes,
such as VSI (vanishing curvature invariants) and CSI (constant curvature invariants) spacetimes.

In section \ref{nullrad}, we study type N vacuum solutions of QG with Ricci tensor containing an additional term
corresponding to null radiation.
In contrast with the results of the previous section, field equations of generic  QG restrict the geometrical properties of the multiple WAND.
It follows that all vacuum solutions of generic QG  with this form of the Ricci tensor belong to the Kundt class of spacetimes.
We then provide examples of such solutions with vanishing and non-vanishing effective cosmological constant $\Lambda$.

\subsubsection{Notation:} 

When appropriate we  work in a  frame {$\bn, \bl, \bm^{(i)}$}
consisting of two null vectors $\bl$, $\bn$ and $n-2$ orthonormal spacelike vectors $\bm^{(i)}$,
\BE
\ell^a \ell_a = n^a n_a = \ell^a m_a^{(i)} = n^a m_a^{(i)} = 0 \, , \qquad
\ell^a n_a = 1 \, , \qquad
m^{(i)a} m_a^{(j)} = \delta_{ij} \, , \label{frame}
\EE
where
$a,b,\dots = 0, \dots n-1$ and $i,j, \dots =2 \dots n-1$ with $n$ being the dimension of the spacetime.

\section{Einstein spacetimes}
\label{sec_einst}

Let us first study Einstein spacetimes 
\begin{equation}
  R_{ab} = \frac{2\Lambda}{n-2} g_{ab}
  \label{Ricci:Einstein}
\end{equation}
as exact solutions to quadratic gravity. 

If we express the Riemann tensor in terms of the Weyl and Ricci tensors and scalar curvature $R = \frac{2n}{n-2}\Lambda$
\begin{equation}
  R_{abcd} = C_{abcd} + \frac{2}{n-2} (g_{a[c} R_{d]b} - g_{b[c} R_{d]a})
  - \frac{2}{(n-1)(n-2)} R g_{a[c} g_{d]b},
  \label{Weyl:definition}
\end{equation}
then for Einstein spacetimes (\ref{Ricci:Einstein}) field equations (\ref{QG:fieldeqns}) reduce to 
\begin{equation}
  \mathcal{B} g_{ab} - \gamma \left( C_a^{\phantom{a}cde} C_{bcde} - \frac{1}{4} g_{ab} C^{cdef} C_{cdef} \right) = 0,
  \label{WNE:fieldeqns}
\end{equation}
where
\begin{equation}
  \mathcal{B} = \frac{\Lambda - \Lambda_0}{2 \kappa}
  + \Lambda^2 \bigg( \frac{(n-4)}{(n-2)^2} (n \alpha + \beta)
  + \frac{(n-3) (n-4)}{(n-2) (n-1)} \gamma \bigg). 
  \label{WNE:fieldeqns:B}
\end{equation}

Note that in the particular case of theories with vanishing Gauss--Bonnet term (i.e.\ $\gamma=0$),
all Einstein spacetimes with $\Lambda$ and $\Lambda_0$ obeying $\mathcal{B}=0$ solve \eqref{WNE:fieldeqns}.
In four dimensions, the Gauss--Bonnet term is purely topological and does not contribute to the field equations
and {moreover the effective cosmological constant $\Lambda$  is equal to  $\Lambda_0$ \eqref{WNE:fieldeqns:B}}.
Thus all four-dimensional Einstein spaces with $\Lambda=\Lambda_0$ solve \eqref{WNE:fieldeqns} as well. 
At the level of the Weyl tensor, this can be seen as a consequence
of the identity $C_a^{\phantom{a}cde} C_{bcde} = \frac{1}{4} g_{ab} C^{cdef} C_{cdef}$ which holds in four dimensions
and which is not valid without additional restrictions for dimensions $n>4$ \cite{Lovelock70}. In the rest of this section
we will study various classes of spacetimes where  $C_a^{\phantom{a}cde} C_{bcde} = \frac{1}{4} g_{ab} C^{cdef} C_{cdef}$
holds in arbitrary dimension due to the vanishing of both terms in the identity.

\subsection{Type N} 

Let us define the following notation
\BE
T_{\{pqrs\}}\equiv\frac{1}{2}(T_{[ab][cd]}+ T_{[cd][ab]}). 
\EE 

The Weyl tensor of type N expressed in the frame \eqref{frame}, where null vector $\bl$ is chosen to coincide with the multiple WAND, admits the form \cite{Coleyetal04} (using notation of \cite{DurkeeetalGHP2010})
\BE
C_{abcd} = 4  \Omega'_{ij}  \ell_{\{ a} {m^i}_b \ell_c {m^j}_{d\}}, \label{WeylN}
\EE
where $\Omega'_{ij}$ is symmetric and traceless.
It follows that for type N spacetimes
\BE
C_a^{\phantom{a}cde} C_{bcde} = C^{cdef} C_{cdef} = 0
\EE    
and \eqref{WNE:fieldeqns} thus reduces to the algebraic constraint 
$\mathcal{B} = 0$ \eqref{WNE:fieldeqns:B}
which, similarly as in the case of (A)dS vacua, prescribes two possible effective cosmological constants $\Lambda$
of the solution for given parameters $\alpha$, $\beta$, $\gamma$, $\kappa$, $\Lambda_0$.\\
{\em Thus in arbitrary dimension all Weyl type N Einstein spacetimes with appropriately chosen
effective cosmological constant $\Lambda$ are exact solutions of quadratic gravity \eqref{QG:fieldeqns}}.\footnote{Note that in the case of the Gauss--Bonnet gravity ($\alpha=\beta=0$) and $\Lambda=0$ this result was already pointed out in \cite{NP2008}. }

Large classes of Einstein spacetimes of type N in $n$ dimensions can be obtained by
warping $(n-1)$-dimensional type N Einstein metric $\d\tilde s^2$, 
\BE
 \d s^2=\frac{1}{f(z)}\d z^2+f(z)\d\tilde s^2 ,  
 \label{ansatz} 
\EE
where
\BE
 f(z)=-\lambda z^2+2dz+b , \qquad \lambda=\frac{2\Lambda}{(n-1)(n-2)} ,
 \label{warpfactor}
\EE
and $b$ and $d$ are constant parameters. Necessary and sufficient condition for $\d s^2$ being an Einstein spacetime is \cite{Brinkmann25}
\BE
 \tilde R=(n-1)(n-2)(\lambda b+d^2),
\label{ricci-n-1}
\EE
where $\tilde R$ is the Ricci scalar of $\d\tilde s^2$. It has been shown \cite{OrtPraPra11warp},
that warping an algebraically special Einstein spacetime leads to an Einstein spacetime of the same principal Weyl type.

Let us briefly overview known type N Einstein spacetimes in higher dimensions. 
The multiple WAND of a type N Einstein spacetime is always geodetic (see \cite{Pravdaetal04} for the
Ricci-flat case) and without loss of generality we  choose an affine parameterization.
Then the optical scalars of the multiple WAND, shear $\sigma^2$, expansion $\theta$ and twist $\omega^2$ are given by \cite{Pravdaetal04,OrtPraPra07}
\BE
 \hspace{-1.3cm} \sigma^2=\ell_{(a;b)}\ell^{(a;b)}-\textstyle{\frac{1}{n-2}}\left(\ell^a_{\;;a}\right)^2 , \qquad \theta=\textstyle{\frac{1}{n-2}}\ell^a_{\;;a} , \qquad 
 \omega^2=\ell_{[a;b]}\ell^{a;b}, 
\EE
respectively. Type N spacetimes can be thus further classified according to the optical properties of the multiple WAND.

The Kundt class of spacetimes for which the multiple geodetic WAND obeys $\theta=0,\ \sigma^2=0,\ \omega^2=0$
admits a metric of the form \cite{ColHerPel06}
\BE
 \d s^2 =2\d u\left[\d v+H(u,v,x^k)\d u+W_{i}(u,v,x^k)\d x^i\right]+ g_{ij}(u,x^k) \d x^i\d x^j.\label{Kundt_gen}
\EE
Einstein Kundt metrics are of principal types II, III and N \cite{OrtPraPra07}. 
For Ricci-flat Kundt metrics of type N and III one can set 
\BE
g_{ij}(u,x^k) = \delta_{ij}, \label{Kundt_vsi}
\EE
with corresponding functions $W_{i}$ and $H$ given in \cite{Coleyetal06}. In addition, the Brinkmann warped
product \cite{OrtPraPra11warp} can be used to generate Einstein type N Kundt metrics with non-vanishing $\Lambda$. 
Examples of such metrics are given in 
\cite{ColHerPel06, ColFusHer09,MalPra11GKS}. 

Expanding ($\theta \not= 0$), non-twisting ($\omega^2=0$) type N Einstein spacetimes: In four dimensions such metrics
are necessarily shear-free due to the Goldberg--Sachs theorem and thus belong to the Robinson--Trautman class.
Metrics for all such spacetimes are known \cite{GarPle81} (see also \cite{GriPodbook} and references therein). 

In contrast for type N Einstein spacetimes in dimensions $n>4$, non-vanishing expansion $\theta \neq 0$ implies $\sigma^2>0$
(see \cite{Pravdaetal04,DurkeeetalGHP2010}). Higher dimensional metrics belonging
to this class can be constructed \cite{OrtPraPra10} by warping four-dimensional Robinson--Trautman type N Einstein spacetimes.

Twisting type N Einstein spacetimes: Very few four-dimensional exact solutions of Einstein gravity within this class are  known.
They include the Ricci-flat Hauser metric \cite{Hauser74} and the Leroy metric \cite{Leroy1970} for negative $\Lambda$ (see also \cite{Stephanibook}).  
Higher dimensional solutions in this class can be  constructed by warping the four-dimensional twisting solutions \cite{OrtPraPra10}.

\subsection{type III}

For type III spacetimes, the Weyl tensor can be expressed as \cite{Coleyetal04}
\BE
C_{abcd} ={{ 8 \Psi'_i  \ell_{\{a} {n}_b \ell
 _c {m^i}_{d\}} + 4 \Psi'_{ijk}  \ell_{\{a} {m^i}_b {m^j}_c {m^k}_{d \}}}}
 + {{4 \Omega'_{ij}  \ell_{\{ a} {m^i}_b \ell_c {m^j}_{d\}}}},
\EE
where 
\BE
\Psi'_{ijk} = -\Psi'_{ikj}, \  \Psi'_{[ijk]}=0, \ \Psi'_i = \Psi'_{kik}. \label{symPsi}
\EE

It follows that $ C^{cdef} C_{cdef} $ vanishes and 
\BE
C_a^{\phantom{a}cde} C_{bcde} = \tilde \Psi  \ell_a \ell_b,
\EE 
where
\BE
\tilde \Psi \equiv  \frac{1}{2} \Psi'_{ijk} \Psi'_{ijk} -  \Psi'_i \Psi'_i. \label{tPsi}
\EE

The trace of \eqref{WNE:fieldeqns} implies $\mathcal{B} = 0$ and therefore 
{\it type III Einstein spacetimes with 
effective cosmological constant $\Lambda$ obeying $\mathcal{B} = 0$ are exact solutions of QG if and only if}
$\tilde \Psi = 0$.

From \eqref{symPsi} it follows that in four dimensions $\tilde \Psi=0$.

Weyl tensor components of Einstein spacetimes obtained by warping seed metrics $\d\tilde s^2$ according to \eqref{ansatz}
and expressed in coordinates
$x^a =(z,x^\mu)$ are \cite{OrtPraPra11warp}
 \BE
 C_{\mu \nu \rho \sigma} = f {\tilde C}_{\mu \nu \rho \sigma}, \ \ \ C_{z \mu \nu \rho} = 0 = C_{z \mu z \nu}.
 \EE
It then follows that the components of $C_a^{\phantom{a}cde} C_{bcde}$ are given by
\BE
C_\mu^{\phantom{\mu}  \nu \rho \sigma} C_{\tau \nu \rho \sigma}  = \frac{1}{f} {\tilde C}_\mu^{\phantom{\mu}  \nu \rho \sigma} {\tilde C}_{\tau \nu \rho \sigma},
\EE 
with all $z$-components being zero. Therefore, $\tilde \Psi$ also vanishes  for all type III Einstein spacetimes obtained by warping four-dimensional type III Einstein spacetimes and these spacetimes are thus also exact solutions of QG. Similarly as in the type N case, we can use seeds $\d\tilde s^2$ with vanishing or non-vanishing expansion and twist. Various classes of  such Einstein spacetimes are given in \cite{OrtPraPra10}.

Let us emphasize that in contrast with the type N case, there exist type III Einstein spacetimes which are {\em not} solutions of QG. For instance, $\tilde \Psi$ is clearly non-vanishing for type III(a) subclass of type III spacetimes characterized by $\Psi'_i =0$ \cite{Coleyetal04}. 
Type III(a) Kundt spacetimes with null radiation given in \cite{Coleyetal06} contain type III(a) Ricci-flat subcases.
An explicit five-dimensional example of such Kundt metric \eqref{Kundt_gen}, \eqref{Kundt_vsi} is given by \cite{OrtPraPra09}
\BEA
 & & W_2=0, \quad W_3=h(u) x^2 x^4,  \quad W_4 = h(u) x^2 x^3,\\
 & & H=H_0 = h(u)^2 \left[ \frac{1}{24} \left( \left(x^3\right)^4+ \left(x^4\right)^4 \right)+ h^0( x^2,x^3,x^4) \right],
\EEA
where $h^0(x^2,x^3,x^4)$ is subject to $\Delta h^0=0$. 
Note that this metric is an example of a type III pp-wave \cite{OrtPraPra09} (pp-waves are defined as spacetimes admitting a covariantly constant null vector). In fact, all type III Ricci-flat pp-waves belong to the type III(a) subclass since the existence of the covariantly constant null vector $\bl$ implies $C_{abcd} \ell^a =0$ and thus $\Psi'_i$ vanishes. Therefore, type III Ricci-flat pp-waves are {\em not} solutions of QG.  

Let us also note that based on the above results it is natural to introduce two new subclasses of the principal type III, namely type III(A) characterized by $\tilde \Psi \not=0$  and type III(B)  characterized by $\tilde \Psi=0$. Obviously type III(a) is a subclass of type III(A).

\subsection{Comparison with other classes of spacetimes}

It is of interest to compare the set of exact solutions of quadratic gravity (QG) with other overlapping classes
of spacetimes, such as {\em pp-waves} (pp-waves of a particular Weyl type will be denoted as ppN, ppIII etc.),
spacetimes with {\em vanishing curvature invariants} (VSI) \cite{Coleyetal04vsi},
spacetimes with {\em constant curvature invariants} (CSI) \cite{ColHerPel06}, {\em Kundt subclass of} CSI (KCSI)
and {\em universal} metrics (U) for which quantum correction is a multiple of the metric \cite{ColGib2008,ColHer2011univ}.
In this discussion we will often consider Einstein or Ricci-flat subsets of these sets.
They will be denoted e.g.\ by ${\mathrm{QG}_{\mathrm{E}}}$ and ${\mathrm{QG}_{\mathrm{RF}}}$.
We consider dimensions $n>4$.\footnote{In four dimensions, all ${\mathrm{pp}_{\mathrm{RF}}}$ are of type N,
which leads to a considerable simplification.}

From the definition of U it follows that U $\subset$ QG. From the results of \cite{Coleyetal04vsi}
it follows that VSI $\subset$ KCSI. pp-waves ${\mathrm{ppN}_{\mathrm{RF}}}$ and ${\mathrm{ppIII}_{\mathrm{RF}}}$
both belong to VSI, but as was shown above ${\mathrm{ppIII}_{\mathrm{RF}}}$ $\cap$ QG is $\varnothing$
and therefore ${\mathrm{VSI}_{\mathrm{RF}}}$ $\not\subset$ QG and ${\mathrm{pp}_{\mathrm{RF}}}$ $\not\subset$ QG.
Note also that in higher dimensions ${\mathrm{ppII}_{\mathrm{RF}}}$ exist \cite{OrtPraPra09}
and thus also ${\mathrm{pp}_{\mathrm{RF}}}$ $\not\subset$ VSI.

Recently it was conjectured in \cite{ColHer2011univ} that U $\subset$ KCSI.
${\mathrm{ppIII}_{\mathrm{RF}}}$ are examples of spacetimes which are KCSI (and VSI) but not U.
Notice however that ${{\mathrm{QG}}_{\mathrm{E}}}$ $\not\subset$ CSI
since examples of ${{\mathrm{QG}}_{\mathrm{E}}}$ metrics with non-vanishing expansion mentioned in this section
have in general non-trivial curvature invariants \cite{OrtPraPra10}.

\section{Type N spacetimes with aligned null radiation}
\label{nullrad}

One may attempt to find a wider class of solutions of \eqref{QG:fieldeqns} considering more general form of the Ricci tensor
than  \eqref{Ricci:Einstein} but still sufficiently simple to considerably simplify \eqref{QG:fieldeqns}.
Thus let us study spacetimes of the Weyl type N with the Ricci tensor of the form
\begin{equation}
  R_{ab} = \frac{2\Lambda}{n-2} g_{ab} + \Phi \ell_a \ell_b, 
  \label{Ricci}
\end{equation}
where $\bl$ coincides with the multiple WAND.
For the Ricci tensor of the form \eqref{Ricci} contracted Bianchi identities $\nabla^a R_{ab} = \frac{1}{2} \nabla_b R$ can be rewritten as
\BE
\left[ \D \Phi + \Phi (n-2) \theta  \right] \ell_a + \Phi \ell_{a;b} \ell^b = 0, \label{Riccinullrad}
\EE
where $\D \equiv \ell^a \nabla_a$. 
This implies that $\bl$ is {\em geodetic} and {without loss of generality we can choose an affine parameterization
of $\bl$ so} that
\begin{equation}
  \D \Phi = - (n-2) \theta \Phi.
  \label{ANR:conservation}
\end{equation}
%Without loss of generality we choose affine parameterization of $\bl$.

Now following the same steps as for Einstein spaces, we express (\ref{QG:fieldeqns}). 
Note that Weyl type N \eqref{WeylN} also implies $C_{abcd} \ell^a$=0 which, together with tracelessness of the Weyl tensor
leads to the vanishing of the terms containing contraction of the Weyl and Ricci tensors.  
The field equations (\ref{QG:fieldeqns}) reduce to 
\begin{equation}
  (\beta \Box + \mathcal{A} ) ( \Phi \ell_a \ell_b )
  -2 \mathcal{B} g_{ab} = 0,
  \label{WNRLNR:fieldeqns:B}
\end{equation}
where
\begin{equation}
  \mathcal{A} = \frac{1}{\kappa}
  + 4 \Lambda \bigg( \frac{n \alpha}{n-2}
  + \frac{\beta}{n-1}
  + \frac{(n-3)(n-4)}{(n-2)(n-1)} \gamma \bigg)
  \label{WNRLNR:fieldeqns:AA}
\end{equation}
and $\mathcal{B}$ is given by \eqref{WNE:fieldeqns:B}.
The trace of (\ref{WNRLNR:fieldeqns:B}) yields $\mathcal{B}=0$ which again determines two possible
effective cosmological constants $\Lambda$ via \eqref{WNE:fieldeqns:B}.
The remaining part of \eqref{WNRLNR:fieldeqns:B} reads
\begin{equation}
  (\beta \Box + \mathcal{A} ) ( \Phi \ell_a \ell_b ) = 0.
  \label{WNRLNR:fieldeqns}
\end{equation}
{Let us briefly comment on the special case $\beta = 0$. Then it follows that  both ${\mathcal A}$ and ${\mathcal B}$ vanish and from \eqref{WNRLNR:fieldeqns:AA} and \eqref{WNE:fieldeqns:B}
one arrives  to
\BE
\frac{\Lambda-2 \Lambda_0}{\kappa}-\frac{8 \Lambda^2 n \alpha}{(n-2)^2}=0.
\EE 
If this constraint on $\Lambda$ admits real solutions, then $\gamma$ is determined from \eqref{WNRLNR:fieldeqns:AA} or \eqref{WNE:fieldeqns:B}.
Therefore for  {\it special} values of the parameters of the theory with $\beta = 0$   equations  \eqref{WNRLNR:fieldeqns} and  ${\mathcal B}=0$ are trivially satisfied and all type N spacetimes with the Ricci tensor of the form \eqref{Ricci} are  exact solutions.
However, we are  interested in solutions of QG with arbitrary parameters $\alpha$, $\beta$, $\gamma$ and such special classes of quadratic gravities are  beyond the scope of this paper. Thus in the rest of this section we assume $\beta \neq 0$.

Contraction of \eqref{WNRLNR:fieldeqns} with vectors $\bl$ and $\bn$ from the frame \eqref{frame} gives
\begin{equation}
\Phi L_{ij} L_{ij} = \Phi [(n-2) \theta^2 + \sigma^2 + \omega^2] = 0,
\end{equation}
where $L_{ij} = \ell_{a;b} m^{(i)a} m^{(i)b}$. 
This implies that $\bl$ is non-expanding ($\theta=0$), shearfree ($\sigma=0$) and non-twisting ($\omega = 0$).
{\em Thus all type N solutions of quadratic gravity with the Ricci tensor of the form \eqref{Ricci} and $\beta \neq 0$ belong to the Kundt class}.

By contracting \eqref{WNRLNR:fieldeqns} with two vectors $\bn$ we obtain the remaining non-trivial component of \eqref{WNRLNR:fieldeqns}
\begin{equation}
  \Box \Phi
  + 4 L_{1i} \delta_i \Phi
  + 2 L_{1i} L_{1i} \Phi 
  + \frac{4 \Lambda \Phi}{n-2} 
  + \mathcal{A} \beta^{-1} \Phi = 0,
  \label{WNRLNR:fieldeqns:comp11b}
\end{equation}
where $L_{1i} \equiv \ell_{a;b} n^{a} m^{(i)b}$.\footnote{The Ricci rotation coefficients, such as the optical matrix $L_{ij}$
and $L_{1i}$ introduced above, appear in the higher dimensional Newman--Penrose formalism \cite{Pravdaetal04,OrtPraPra07,Coleyetal04vsi} (see also  \cite{DurkeeetalGHP2010}).
In the derivation of \eqref{WNRLNR:fieldeqns:comp11b} we have used some of the Ricci equations of \cite{OrtPraPra07}
when appropriate. 
However, note  that for our purposes the use of the formalism is not essential and one can  work with equation
\eqref{WNRLNR:fieldeqns} instead of \eqref{WNRLNR:fieldeqns:comp11b}. In addition we also used the fact that for the Kundt metrics in the canonical form \eqref{Kundt_gen},
 $L_{1i}=L_{i1}$.}

\subsection{Explicit solutions}
\subsubsection{Case $\Lambda=0$}
Type N Kundt metrics with null radiation and vanishing cosmological constant
admit the form \eqref{Kundt_gen}, \eqref{Kundt_vsi}
%\begin{equation}
%  \d s^2 = 2 \d u [ \d v + H \d u + W_i \d x^i ] + \delta_{ij} \d x^i \d x^j,
%  \label{VSI:metric}
%\end{equation}
with the frame vectors  given by \cite{Coleyetal06}
\begin{eqnarray}
  \bl &=& \d u \, , \quad
  \bn = \d v + H \d u + W_i \d x^i \, , \quad
  \bm^{(i)} = \d x^i \, , \\
  \bl &=& \partial_v \, , \quad
  \bn = \partial_u - H \partial_v \, , \quad
  \bm_{(i)} = \partial_i - W_i \partial_v \, .
  \label{VSI:frame}
\end{eqnarray}
These metrics split into two subclasses with vanishing ($\varepsilon=0$) or non-vanishing ($\varepsilon=1$)
quantity $L_{1i} L_{1i}$.

The type N condition on the Weyl tensor and the form \eqref{Ricci} of the Ricci tensor impose
the following constraints on the undetermined metric functions \cite{Coleyetal06} in the case $\varepsilon=0$
\begin{equation}
  W_2 = 0, \qquad
  W_{\tilde{\imath}} = x^2 C_{\tilde{\imath}}(u) + x^{\tilde{\jmath}} B_{\tilde{\jmath}\tilde{\imath}}(u), \qquad
  H = H^0(u,x^i),
\end{equation}
\begin{equation}
  \Delta H^0 - \frac{1}{2} \sum C^2_{\tilde{\imath}} - 2 \sum_{\tilde{\imath}<\tilde{\jmath}} B^2_{\tilde{\imath}\tilde{\jmath}} + \Phi = 0,
  \label{VSI:EFEs:0}
\end{equation}
and in the case $\varepsilon=1$ 
\begin{equation}
  W_2 = - \frac{{{2}} v}{x^2}, \qquad
  W_{\tilde{\imath}} = C_{\tilde{\imath}}(u) + x^{\tilde{\jmath}} B_{\tilde{\jmath}\tilde{\imath}}(u), \qquad
  H = \frac{v^2}{2(x^2)^2} + H^0(u,x^i),
\end{equation}
\begin{equation}
  x^2 \Delta \left(\frac{H^0}{x^2}\right) - \frac{1}{(x^2)^2} \sum W^2_{\tilde{\imath}} - 2 \sum_{\tilde{\imath}<\tilde{\jmath}} B^2_{\tilde{\imath}\tilde{\jmath}} + \Phi = 0,
  \label{VSI:EFEs:1}
\end{equation}
where $B_{[\tilde{\imath}\tilde{\jmath}]}=0$ in both cases and $\tilde{\imath},\ \tilde{\jmath}, \dots =3 \dots n-1$.

For these metrics \eqref{WNRLNR:fieldeqns} reduces to
\begin{equation}
  \Phi_{,ii} - \frac{2 \varepsilon}{x^2} \Phi_{,2} + \frac{2 \varepsilon}{(x^2)^2} \Phi + (\kappa\beta)^{-1} \Phi = 0,
\end{equation}
or
\begin{align}
  \Delta \Phi + (\kappa \beta)^{-1} \Phi = 0 \qquad (\varepsilon=0)
  \label{VSI:Phi:0}, \\
  x^2 \Delta \left( \frac{\Phi}{x^2} \right) + (\kappa \beta)^{-1} \Phi = 0 \qquad (\varepsilon=1)
  \label{VSI:Phi:1}.
\end{align}

Using \eqref{VSI:Phi:0}, \eqref{VSI:Phi:1}, we can rewrite \eqref{VSI:EFEs:0}, \eqref{VSI:EFEs:1} as
\begin{equation}
  \Delta H^0_{\mathrm{vac}} - \frac{1}{2} \sum C^2_{\tilde{\imath}} - 2 \sum_{\tilde{\imath}<\tilde{\jmath}} B^2_{\tilde{\imath}\tilde{\jmath}} = 0
  \label{VSI:vacuumEFEs:0},
\end{equation}
\begin{equation}
  x^2 \Delta \left(\frac{H^0_{\mathrm{vac}}}{x^2}\right) - \frac{1}{(x^2)^2} \sum W^2_{\tilde{\imath}} - 2 \sum_{\tilde{\imath}<\tilde{\jmath}} B^2_{\tilde{\imath}\tilde{\jmath}} = 0 \, ,
  \label{VSI:vacuumEFEs:1}
\end{equation}
respectively, with $H^0_{\mathrm{vac}} = H^0 - \kappa \beta \Phi$ corresponding to a vacuum solution of the Einstein gravity
(\eqref{VSI:EFEs:0}, \eqref{VSI:EFEs:1} with $\Phi=0$). 
In other words, one may take an arbitrary \emph{vacuum $(\Phi=0)$ solution of the Einstein field equations}
\eqref{VSI:vacuumEFEs:0} or \eqref{VSI:vacuumEFEs:1} and independently find $\Phi$ solving corresponding equation
\eqref{VSI:Phi:0} or \eqref{VSI:Phi:1} and arrive to a \emph{vacuum  solution of quadratic gravity}
with $\Phi \not=0$, $H^0 = H^0_{\mathrm{vac}} + \kappa \beta \Phi$ and $W_i$ unchanged.

Note that due to \eqref{WNE:fieldeqns:B} the assumption $\Lambda = 0$ implies $\Lambda_0 = 0$  and therefore we are not able to satisfy criticality condition \cite{Deser2011} by tuning the remaining parameters $\alpha$, $\beta$ and $\gamma$.}

\subsubsection{Case $\Lambda \not=0$}

One may perform a similar procedure in the case of non-vanishing effective cosmological constant $\Lambda$.
As an example of type N Kundt metric with non-vanishing $\Lambda$ we take the $n$-dimensional
Siklos metric \cite{ChamGib00}
\begin{equation}
  \d s^2 = \frac{1}{-\lambda z^2} \left( 2 \d u \d v  + 2 H(u, x^k) \, \d u^2 + \delta_{ij} \d x^i \d x^j \right) ,
  \label{Siklos}
\end{equation}
where
\begin{equation}
  \lambda = \frac{2 \Lambda}{(n-1)(n-2)}, \qquad z = x^n.
  \label{}
\end{equation}
The non-expanding geodetic null congruence is  $\bk = k_a \d x^a = \d u$.  
The condition \eqref{Ricci} on the Ricci tensor reads
\begin{equation}
  \Delta H - \frac{n-2}{z} H_{,z} = \Phi
  \label{Siklos:EFEs}
\end{equation}
and \eqref{WNRLNR:fieldeqns} can be expressed as
\begin{equation}
  \Delta\!\left( -\lambda z^2 \Phi \right) - \frac{n-2}{z} \left( -\lambda z^2 \Phi \right)_{,z}
  - \frac{\mathcal{C}}{z^2} ( -\lambda z^2 \Phi ) =0,
  \label{Siklos:Phi}
\end{equation}
where we defined
\begin{equation}
  \mathcal{C} \equiv \frac{2 \lambda + \mathcal{A} \beta^{-1}}{\lambda}
  = \frac{2}{\beta} \left( \frac{1}{2 \lambda \kappa} + (n-1)(n \alpha + \beta) + (n-3)(n-4) \gamma \right).
  \label{}
\end{equation}
Using \eqref{Siklos:Phi}, and denoting $H^{\rm vac} = H - \mathcal{C}^{-1} z^2 \Phi$ 
we can rewrite \eqref{Siklos:EFEs} as  
\begin{equation}
  \Delta H^{\rm vac} - \frac{n-2}{z} H^{\rm vac}_{,z} = 0.
  \label{Siklos:EFEs:vacuum}
\end{equation}
Therefore we can take an arbitrary Einstein metric \eqref{Siklos} solving
\eqref{Siklos:EFEs:vacuum},\footnote{Large class of such vacuum solutions of Einstein gravity
can be found in \cite{ChamGib00}.} and find a solution $\Phi$ of \eqref{Siklos:Phi}.
Then the metric \eqref{Siklos} with $H = H^{\rm vac} + \mathcal{C}^{-1} z^2 \Phi$ and $\Phi$ solve
the QG field equations \eqref{QG:fieldeqns}.\footnote{Note that this method for generating solutions of QG cannot be used
for critical points of quadratic gravities \cite{Deser2011} since in such case $\mathcal{C}=0.$
%{\bf [and one can no longer solve $H$ and $\Phi$ from \eqref{Siklos:EFEs}, \eqref{Siklos:Phi} independently.]}
}

\begin{acknowledgments}
We thank Marcello Ortaggio and Alena Pravdov\'a for helpful discussions.  
V.\ P.\ has been supported by research plan no AV0Z10190503 and research grant GA\v CR
P203/10/0749. T.\ M.\ is supported by the project SVV 263301 of the Charles University in Prague.
Some of the equations in this paper were derived or checked using the computer algebra package Cadabra
\cite{Peeters2006,Peeters2007}
\end{acknowledgments}

\end{document}